\documentclass[aps,prb,amsmath,amssymb,footinbib,showpacs,twocolumn,superscriptaddress]{revtex4-1}
\usepackage{amsmath}
\usepackage{amssymb}
\usepackage{amsthm}
\usepackage{setspace}
\usepackage{graphicx}
\usepackage{braket}
\usepackage{mathrsfs}
\usepackage{physics}
\usepackage{float}
\usepackage[colorlinks = true,linkcolor = red,urlcolor  = blue,citecolor = blue,anchorcolor = blue]{hyperref}
\usepackage[utf8]{inputenc}
\usepackage[english]{babel}

\begin{document}
\title{Quasiparticle gaps  in multiprobe Majorana nanowires}
\author{Yingyi Huang}
\affiliation{
Condensed Matter Theory Center and Joint Quantum Institute, Department of Physics, University of Maryland, College Park, Maryland 20742, USA}
\affiliation{
State Key Laboratory of Optoelectronic Materials and Technologies, School of Physics, Sun Yat-sen University, Guangzhou 510275, China}
\author{Jay D. Sau}
\affiliation{
Condensed Matter Theory Center and Joint Quantum Institute, Department of Physics, University of Maryland, College Park, Maryland 20742, USA}
\author{Tudor D. Stanescu}
\affiliation{
Condensed Matter Theory Center and Joint Quantum Institute, Department of Physics, University of Maryland, College Park, Maryland 20742, USA}
\affiliation{
Department of Physics and Astronomy, West Virginia University, Morgantown, West Virginia 26506, USA}
\author{S. Das Sarma}
\affiliation{
Condensed Matter Theory Center and Joint Quantum Institute, Department of Physics, University of Maryland, College Park, Maryland 20742, USA}

\begin{abstract}
We theoretically study a spin-orbit-coupled nanowire proximitized by a superconductor in the presence of an externally applied Zeeman field (``Majorana nanowire") with zero-energy Majorana bound states localized at the two ends of the wire when the Zeeman spin splitting is large enough for the system to enter the topological phase. The specific physics of interest in the current work is the effect of having several tunnel probes attached to the wire along its length. Such tunnel probes should allow, as a matter of principle,  one to observe both the predicted bulk superconducting gap closing and opening associated with the topological quantum phase transition as well as the Majorana bound states at the wire ends showing up as zero-bias conductance peaks, depending on which probes are used for the tunneling spectroscopy measurement. Because of the possible invasive nature of the tunnel probes, producing local potential fluctuations in the nanowire, we find the physical situation to be quite complex. In particular, depending on the details of the tunnel barrier operational at the probes, the Majorana nanowire could manifest additional low-energy Andreev bound states which will manifest their own almost-zero-bias peaks, complicating the interpretation of the tunneling data in multiprobe Majorana nanowires. We use two complementary microscopic models to simulate the probes, finding that the tunneling conductance spectrum depends rather sensitively on the details of the tunnel barriers at the probes, but in some situations it should be possible to observe the Majorana bound-state-induced zero-bias conductance peak at the wire ends along with the gap closing and opening features associated with the bulk topological quantum phase transition in the multiprobe Majorana nanowires. Our detailed numerical results indicate that such a system should also be capable of directly manifesting the nonlocal conductance correlations arising from Majorana bound states at the two ends of the nanowire. We apply our general analysis to simulate a recent multiprobe nanowire experiment commenting on the nature of the quasiparticle gaps likely controlling the experimental observations.
\end{abstract}
\date{\rm\today}
\maketitle
\section{introduction}\label{sec:introduction}
Majorana zero modes (MZMs) localized at the ends of one-dimensional (1D) topological superconductors have become the object of a fervent experimental research~\cite{Nayak2008Non-Abelian, DasSarma2015Majorana,Alicea2012New,Elliott2015Colloquium,Stanescu2013Majorana,Leijnse2012Introduction,Beenakker2013Search,Lutchyn2017Realizing,jiang2013non,sato2016majorana,Sato2016Topological,Aguado2017Majorana} following the concrete prescription for the realization of a topological superconducting phase using spin-orbit-coupled semiconductor (SM) nanowires proximity-coupled to $s$-wave superconductors (SCs) in the presence of an external magnetic field~\cite{Sau2010Generic,Lutchyn2010Majorana,Oreg2010Helical,Sau2010Non}. Increasing the magnetic field drives the hybrid system through a topological quantum phase transition (TQPT), from a trivial to a topological SC phase, with the bulk quasiparticle gap vanishing at the TQPT and reopening as a topological gap that hosts a pair of midgap MZMs localized at the opposite ends of the system~\cite{Sau2010Generic,Lutchyn2010Majorana,Oreg2010Helical,Sau2010Non}.  The primary method for detecting the presence of  MZMs is measuring the charge tunneling current into the edge of the 1D superconductor, which is predicted to result in a zero-bias conductance peak (ZBCP) with a quantized height ($2e^2/h$) at low temperature~\cite{Sengupta2001Midgap, Law2009Majorana,Lin2012Zero,Prada2012Transport}. Nonquantized ZBCPs have been observed in many InAs- and InSb-based hybrid systems~\cite{Mourik2012Signatures,Das2012Zero,Deng2012Anomalous,Churchill2013Superconductor,Finck2013Anomalous,Albrecht2016Exponential,Chen2016Experimental,Zhang2016Ballistic,Deng2016Majorana} over the past few years. More recently, the observation of a quantized ZBCP has also been reported~\cite{Zhang2017Quantized}. Unfortunately, however, the observation of a ZBCP, just by itself, is not a decisive hallmark of the topologically protected MZM. It has been shown theoretically that low-energy Andreev bound states (ABSs) mimicking the phenomenology of MZMs can emerge in the trivial SC phase (i.e., before the TQPT) in the presence of inhomogeneous effective potentials~\cite{Kells2012Near,Brouwer2011Topological,Stanescu2014Nonlocality,Moore2016Majorana} or in systems with quantum dots attached to the end of the 1D superconductor~\cite{Prada2012Transport,Liu2017Andreev,Moore2018Quantized}. In turn, the presence of these low-energy ABSs may generate robust ZBCPs~\cite{Kells2012Near,Prada2012Transport,Fleckenstein2018Decaying,Liu2017Andreev,Moore2016Majorana,Setiawan2017Electron} and even quantized ZBCPs~\cite{Moore2018Quantized,Vuik2018Reproducing} that are indistinguishable from Majorana-induced ZBCPs. In fact, these almost-zero-energy ABSs in Majorana nanowires can be thought of as ``quasi-MZMs", since they are nothing but multiple (at least, two) even-numbered MZMs spatially located not very far from each other.  By contrast, the real MZMs above the TQPT are located at the wire ends spatially far separated from each other.

The most straightforward generalization of the standard end-of-wire tunneling measurement that can shed light on the nature of the quasiparticles responsible for the observed ZBCPs involves a multiprobe tunneling experiment that provides simultaneous tunneling measurements at different locations along the wire. One proposal suggests performing two tunneling measurements at the ends of the wire~\cite{DasSarma2012Splitting}: the presence of MZMs is signaled by (nonlocal) correlations of the observed features (e.g., finite-size-induced energy splittings and fields associated with the emergence of ZBCPs), while the absence of such correlations would  suggest trivial ABS-induced  ZBCPs. Another possibility is to perform  simultaneous end-of-wire and bulk measurements~\cite{Stanescu2012To}: the emergence at the end of the wire of a Majorana-induced ZBCP above a certain critical field corresponding to the TQPT should be accompanied by the opening of a bulk topological gap, while the emergence of the ZBCP before the closing and reopening of the bulk gap signals its ABS origin. 

Recently, the experimental observation of a concomitant opening of a bulk gap with an end-of-wire ZBCP has been reported in Ref.~[\onlinecite{Grivnin2018concomitant}] by Grivnin \textit{et al}. While this observation is consistent with a Majorana-induced ZBCP (and at some level, with the predictions of Ref.~[\onlinecite{Stanescu2012To}]),  some important questions still remain, first and foremost regarding the nature of the low-energy state responsible for the gap closing and reopening measured by the bulk probe. As demonstrated in Ref.~[\onlinecite{Stanescu2018Building}], attaching a contact to the bulk of a hybrid system may result in a local perturbation that induces a low-energy ABS localized near the contact region. Consequently, rather than measuring the lowest energy bulk state which is responsible for the reopening of the topological gap above the TQPT, the bulk contact probes a contact-induced low-energy ABS that, unlike the delocalized bulk states, is strongly coupled to the lead (due to its localized nature). The details obviously depend on the nature of the probes, but one has little experimental control over such probes, and fabricating probes which are robust enough to measure the tunneling current but at the same time delicate enough not to be invasive is a tall order.  

In this work we investigate the multiprobe Majorana hybrid structure in light of the recent interesting experimental results~\cite{Grivnin2018concomitant}, focusing on a specific question: Is the opening of a gap (measured by the bulk probe) concomitant with the emergence of a ZBCP (at the end of the wire) the decisive hallmark of a topological SC phase (and its associated MZMs)? To address this question, we study numerically two different (but related) tight-binding models. The first approach assumes the formation of a Schottky barrier~\cite{Schottky1939halbleitertheorie} at the junction between the normal leads and the proximitized SM nanowire, which is modeled as a weak link, in addition to a finite electrostatic barrier. Note that this approach is  similar in spirit to the treatment of SM-SM and SM-SC junctions in Ref.~[\onlinecite{DasSarma2016How}]. The second approach takes into account the fact that the presence of a metallic lead may produce a stronger (or, more generally, a different) screening in the contact region. We incorporate this effect  as a position-dependent electrostatic potential that takes different values  in the contact and contact-free regions.  We emphasize that the detailed nature of the tunnel probe along the nanowire in Ref.~[\onlinecite{Grivnin2018concomitant}] is unknown, and hence our two models are two complementary minimal physical descriptions of the system. The real system is likely to be much more complex, but there is not enough information to simulate the precise real system, and in any case, the reality is most likely a combination of our two minimal models in some complicated manner. 

To investigate the low-energy local properties at different locations along the wire in the presence of the contact-induced perturbations, we calculate the local density of states (LDOS) in the contact regions (using a Green's function formalism) as well as the charge tunneling conductance associated with each lead (using standard scattering matrix formalism). We find that, within a significant range of parameters, the lowest-energy Bogoliubov-de Gennes (BdG) state in the topological regime is a Majorana mode, while the second-lowest-energy excitation is an Andreev bound state localized near the bulk probe, rather that a genuine bulk state. Nonetheless,  in the weak-perturbation regime, the contact-induced ABS crosses zero energy close to the TQPT, producing a feature in the tunneling conductance that mimics the closing and reopening of the bulk quasiparticle gap at the TQPT. If the setup contains multiple bulk probes, the measured ``bulk'' gaps are likely to be different, as the corresponding probe-induced ABSs are sensitive to the local conditions, which may be slightly different for different contacts. This is in contrast with the behavior of a true topological bulk gap, which must have the same value throughout the entire system.  Reference~[\onlinecite{Grivnin2018concomitant}] reports different values of the quasiparticle gaps measured by different bulk probes, suggesting a most likely probe-induced ABS nature of these features. Furthermore, we find that for certain values of the chemical potential the probe-induced ABS may ``stick'' to zero energy over a certain range of Zeeman fields before reacquiring a finite gap. This type of behavior generates ZBCP-like features {\em in the bulk}, an element that is also present in the recent experiment~\cite{Grivnin2018concomitant}. Note that, in a finite system, the bulk quasiparticle gap never closes completely, but has a minimum at some ``critical'' value of the applied field, which is inconsistent with the emergence of  ZBCP-like features. Based on our numerical results, we conclude that the ``bulk'' features observed in the recent experiment are likely generated by contact-induced ABSs that approach zero energy in the vicinity of the TQPT and become gapped at larger magnetic fields. These ``bulk'' features approach the edge of the true topological (bulk) gap in the limit of weakly invasive probes. We provide concrete suggestions on how to achieve this regime in  multiprobe Majorana nanowires.

The rest of this paper is organized as follows: In Sec. ~\ref{sec:model1}, we consider a minimal model of the multiprobe hybrid structure and study it numerically, focusing on the local density of states and the tunneling conductance. A more detailed model that incorporates the possible perturbation of the effective electrostatic potential by the probes is considered in Sec.~\ref{sec:model2}. Our results are summarized in Sec.~\ref{sec:conclusion}, together with a specific proposal for improving the experimental setup in light of these results.

\section{Contact barrier model}\label{sec:model1}
\begin{figure}
\raggedleft
\includegraphics[width=1.0\columnwidth]{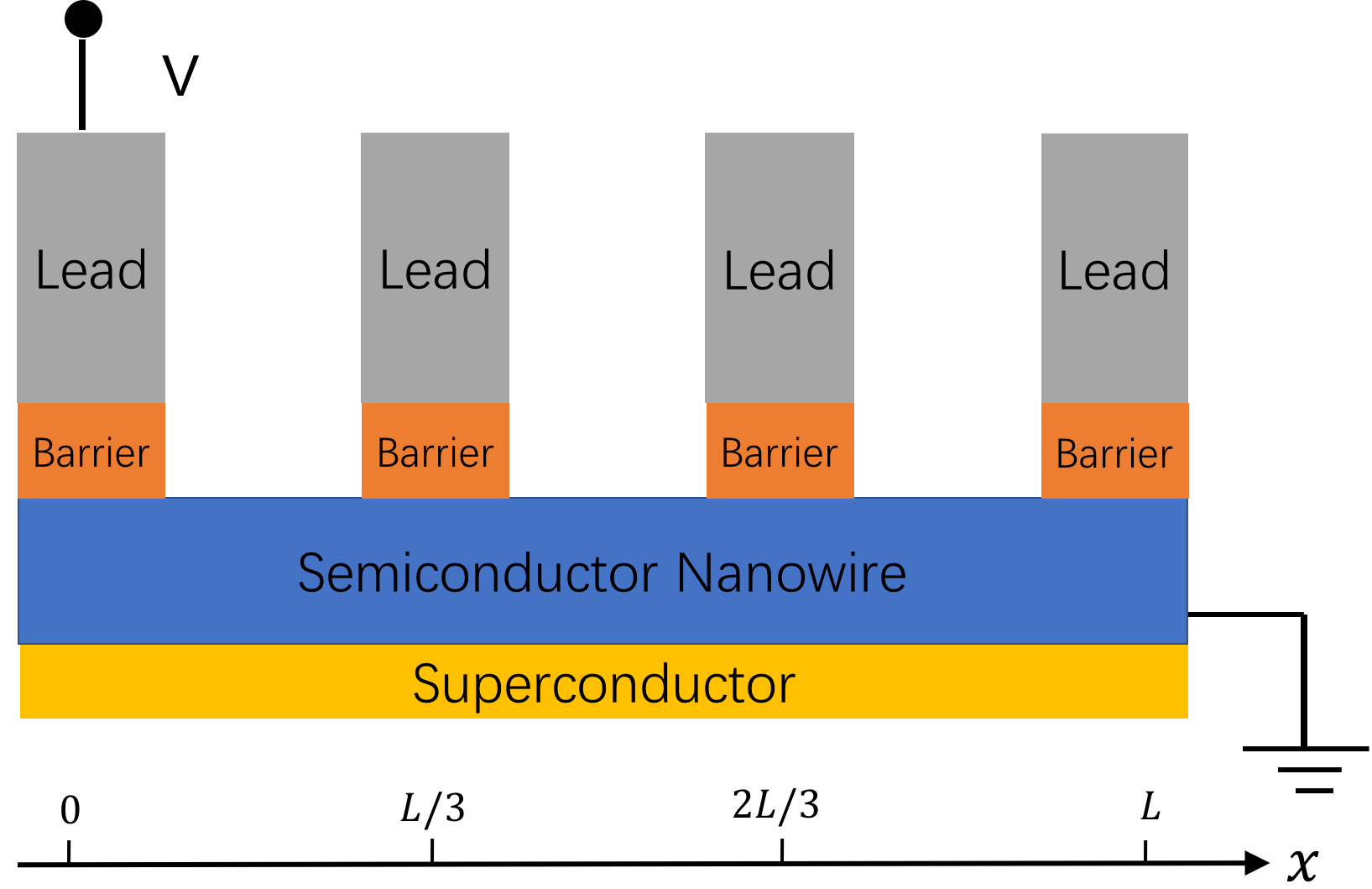}
\caption{(Color online) Schematic diagram for a device consisting of a superconductor-semiconductor nanowire, four normal metal leads, and four potential barriers. As parts of the system, the potential barriers in a length $L_b$ are uncovered by superconductor and subject to a square potential $V_{b}$.
 The tunneling conductance $G(x)$ for position $x=0,L/3,2L/3,L$ is obtained by applying a voltage bias $V$ to the lead at $x$ and measuring the tunneling current through the nanowire.  }
\label{fig:schematic1}
\end{figure}

We start by considering a microscopic model for a Rashba semiconductor nanowire in proximity to an $s$-wave superconductor and in contact with four leads (shown in Fig.~\ref{fig:schematic1}).  At the interface between the nanowire and any one of the four leads, there is always a potential barrier, no matter whether the leads are actively tunneling to the wire or not. The four potential barriers are localized along the nanowire in $x$ direction, i.e., two of the barriers are at the wire end ($x=0,L$) and the other two are in the bulk of the wire ($x=L/3,2L/3$). This set up is a qualitative minimal model for the multiprobe nanowire of Ref.~[\onlinecite{Grivnin2018concomitant}].

To describe the coupling terms in a convenient way, we write down the Hamiltonian in a discretized tight-binding form. The total Hamiltonian has three terms,
\begin{equation}
H=H_{NW}+H_{NW-barr}+H_{barr},
\end{equation}
where $H_{NW}$ and $H_{barr}$ are the Hamiltonians of the Majorana nanowire and of the barriers, respectively, and $H_{NW-barr}$ describes the coupling between the nanowire and the barriers, i.e., the link.

The generic noninteracting low-energy effective Hamiltonian for the nanowire is~\cite{Sau2010Generic, Lutchyn2010Majorana, Oreg2010Helical}
\begin{align}\label{eq:HNW}
H_{NW}&=\sum_{j\in \mathcal{L}_{NW}} \Big[-tc^\dag_{j+\delta_x}c_j+(-\mu+2t)c^\dag_{j}c_{j} \\ \nonumber
&+i\alpha(c^\dag_{j+\delta_x}\hat{\sigma}_yc_j-\text{H.c.})+V_Zc^\dag_{j}\hat{\sigma}_xc_j+\Sigma(\omega,V_Z)c^\dag_{j}c_{j}\Big],  \nonumber
\end{align}
where $c_i^\dag=(c_{i\uparrow}^\dag,c_{i\downarrow}^\dag)$ are spinors, with $c^\dag_{i\sigma}$ being the electron creation operator for spin $\sigma$.  Here, the summation is over the sites in the nanowire $\mathcal{L}_{NW}$. $\sigma_\mu$ ($\tau_\mu$) are Pauli matrices in spin (particle-hole) space, $\delta_x$ gives the nearest neighbors along $x$-direction, and $\mu$ is the chemical potential. The hopping amplitude and spin-orbit coupling (SOC) are given as $t=\frac{1}{2m^*a^2}$ and $\alpha=\frac{\alpha_R}{2a}$ by the effective mass $m^*$ and Rashba SOC strength $\alpha_R$ with a lattice constant $a$. The Zeeman spin splitting energy $V_Z$ has a relation with magnetic field $B$ as $V_Z=g^*\mu_B V$, with $g^*$ being Land\'e $g$ factor and $\mu_B$ the Bohr magneton.

The proximity SC effect~\cite{Stanescu2010Proximity,DasSarma2012Splitting,Stanescu2017Proximity} is captured by a self-energy term at energy $\omega$ as
\begin{equation}
\Sigma(\omega,V_Z)=-\lambda\frac{\omega\tau_0+\Delta(V_Z)\tau_x}{\sqrt{\Delta(V_Z)^2-\omega^2}},
\end{equation}
where $\lambda$ is the effective coupling at the superconductor-semiconductor interface. At low energy, the self-energy term can be approximated by $-\lambda\tau_x$. Accordingly, the critical Zeeman field for the topological quantum phase transition between trivial and topological superconducting phases can be described by $V_{Zc}=\sqrt{\mu^2+\lambda^2}$.

As a common experimental scenario~\cite{Deng2016Majorana,Nichele2017Scaling}, the bulk SC gap is chosen to be a $V_Z$-dependent form as
\begin{equation}
\Delta(V_Z)=\Delta_0\sqrt{1-(V_Z/V_Z^*)^2},
\end{equation}
with the bulk gap being suppressed by increasing Zeeman field up to $V_Z^*$. The induced gap of the system can be described by $\Delta_{ind}=\lambda\Delta(V_Z)/[\lambda+\Delta(V_Z)]$.

In this model, the coupling strength between the nanowire and the leads is determined by two factors.  The first is the height of the barrier in $H_{barr}$ [Eq.~\eqref{eq:HBarrier}], and the second is the coupling strength of the weak link in $H_{NW-barr}$ [Eq.~\eqref{eq:HNW-Barrier}].

The barrier by definition is subjected to a finite electrostatic potential and without induced SC, thus the barrier Hamiltonian is
\begin{align}\label{eq:HBarrier}
H_{barr}&=\sum_{j\in \mathcal{L}_{barr}}\Big[-t_bc^\dag_{j+\delta_x}c_j+(-\mu+2t_b+V_b)c^\dag_{j}c_{j}\\ \nonumber
&+i\alpha(c^\dag_{j+\delta_x}\hat{\sigma}_yc_j-\text{H.c.})+V_Zc^\dag_{j}\hat{\sigma}_xc_j\Big],  \nonumber
\end{align}
with the summation over the sites in the four barriers $\mathcal{L}_{barr}$. Here, the hopping amplitude of the barriers $t_b$ is smaller than the nanowire hopping $t$, since the probe is a normal metal in the experiment.

The coupling Hamiltonian between the nanowire and the four barriers is
\begin{equation}\label{eq:HNW-Barrier}
H_{NW-barr}=\sum_{j\in \mathcal{P}_{NW}}\sum_{j'\in \mathcal{P}_{barr}}\Big[-t'c^\dag_{j}c_{j'}+i\alpha'(c^\dag_{j}\hat{\sigma}_yc_{j'}-\text{H.c.})\Big],
\end{equation}
where $\mathcal{P}_{NW}$ and $\mathcal{P}_{barr}$ are the sites in the nanowire, and the barriers near their interface,
and $t'$ and $\alpha'$ correspond to hopping and spin-orbit coupling in the links between the nanowire and the barriers, respectively. For weaker link couplings $t'$ and $\alpha'$, the tunnel barrier is higher~\cite{DasSarma2016How}.

The local density of states at energy $\omega$ at a different position $j$ of the nanowire can be calculated from Green's function $G(\omega)=[\omega-H]^{-1}$ as
\begin{equation}
\rho(\omega,j)=-\frac{1}{\pi}\operatorname{Im}{\operatorname{Tr}[G(\omega+i\eta,j,j)]},
\end{equation}
with $\eta\rightarrow0^+$.
Here, $\operatorname{Im}{...}$ represents the imaginary part and $\operatorname{Tr}[...]$ represents the trace over spin and particle-hole degrees of freedom.
To increase the visibility of LDOS in the numerical calculation, $\eta$ should be chosen to be higher.

\begin{figure*}[t!]
\raggedleft
\includegraphics[width=1.0\textwidth]{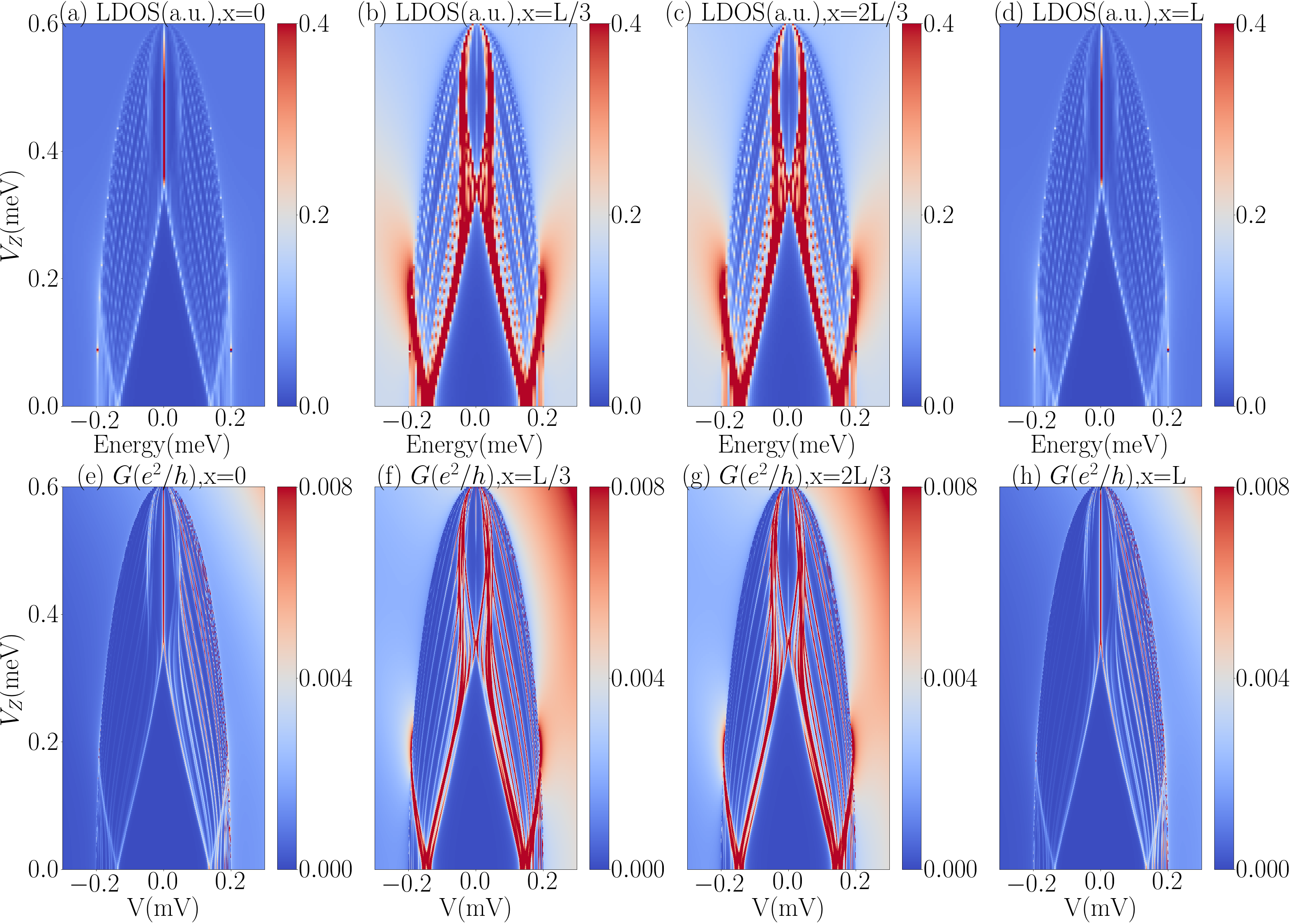}
\caption{(Color online) Local density of states (upper panels) and differential conductance (lower panels) as a function of Zeeman field at positions $x=0, L/3, 2L/3, L$ from left to right along the nanowire attached to four barriers (shown in Fig.~\ref{fig:schematic1}).
Most parameters are chosen to be consistent with the Grivnin \textit{et al}. experiment~\cite{Grivnin2018concomitant}:  $L=2.5~\mu$m, $L_b=0.1~\mu$m, $\mu=0.15$~meV, $\Delta_0=0.2$~meV, $\lambda=0.3$~meV, $V_{Z}^*=0.6$~meV, $V_b=1$~meV, $\eta=0.003$~meV, $\Gamma=0.003$~meV, $a=0.01~\mu$m, $t=16.7$~meV, $\alpha=0.5$~meV, $t_b=10$~meV, $t'=5$~meV, and $\alpha'=0.25$~meV.
}
\label{fig:LG11}
\end{figure*}
\begin{figure}
\includegraphics[width=0.8\columnwidth]{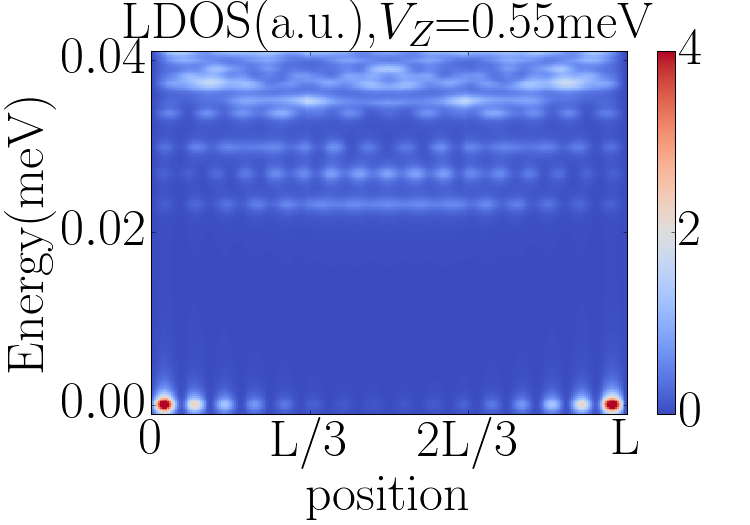}
\caption{(Color online) Spatial profiles of LDOS of the nanowire shown in Fig.~\ref{fig:schematic1} at $V_Z=0.55$~meV in the topological region ($V_Z>V_{Zc}=0.34$~meV). The imaginary term is chosen to be $\eta=0.001$~meV, and the other parameters are the same as those of \ref{fig:LG11}.}
\label{fig:LDOSx11}.
\end{figure}

In order to obtain the conductance, we attach a semi-infinite long lead to the nanowire through one of the barriers.
 The lead Hamiltonian is
\begin{align}\label{eq:Hlead}
H_{lead}&=\sum_{j} \Big[-t_bc^\dag_{j+\delta_x}c_j+(-\mu_{lead}+2t_b)c^\dag_{j}c_{j} \\ \nonumber
&+i\alpha(c^\dag_{j+\delta_x}\hat{\sigma}_yc_j-\text{H.c.})+V_Zc^\dag_{j}\hat{\sigma}_xc_j\Big],\nonumber
\end{align}
with the chemical potential of the lead being $\mu_{lead}$.  Applying a voltage bias $V$ and measuring the tunneling current $I$, the tunneling differential conductance ($G=dI/dV$) can be expressed in terms of the elements of the corresponding $S$ matrix~\cite{Blonder1982Transition}.
The numerical calculation of the $S$ matrix is implemented by a Python package KWANT~\cite{kwant}.  For conductance data smoothening, an infinitesimal on-site imaginary term $i\Gamma$ has been added into the nanowire Hamiltonian [Eq.~\eqref{eq:HNW}], which will lead to particle-hole asymmetry of the conductance at finite energies~\cite{DasSarma2016How,Liu2017Role}.

For the results presented in this section, we choose two sets of parameter values (i.e., with finite and zero chemical potential) for the tight-binding model. The results are presented in Secs.~\ref{sec:finitemu} and \ref{sec:zeromu}, respectively. Since some of the applicable parameters for the Grivnin \textit{et al}. experimental system are empirically estimated~\cite{Grivnin2018concomitant} (for example, the induced gap is $\Delta_{ind}\sim[0.12,0.15]$~meV, the bulk gap closes at $B\approx0.9$~T, the bulk superconductivity  is destroyed at $B\approx1.6$~T,  and the length of nanowire is $L\approx2.5~\mu$m), we estimate the parameters and vary them in a small range. Note that the parameters are spatially uniform along the wire in spite of the coupling to the barriers. All the results are obtained at zero temperature. We include no disorder or chemical potential fluctuation in the nanowire itself, unlike, e.g., Ref.~[\onlinecite{Liu2017Andreev}], and hence the nanowire has no quasi-MZM or zero-energy ABS without the leads.

\begin{figure*}[t!]
\includegraphics[width=1.0\textwidth]{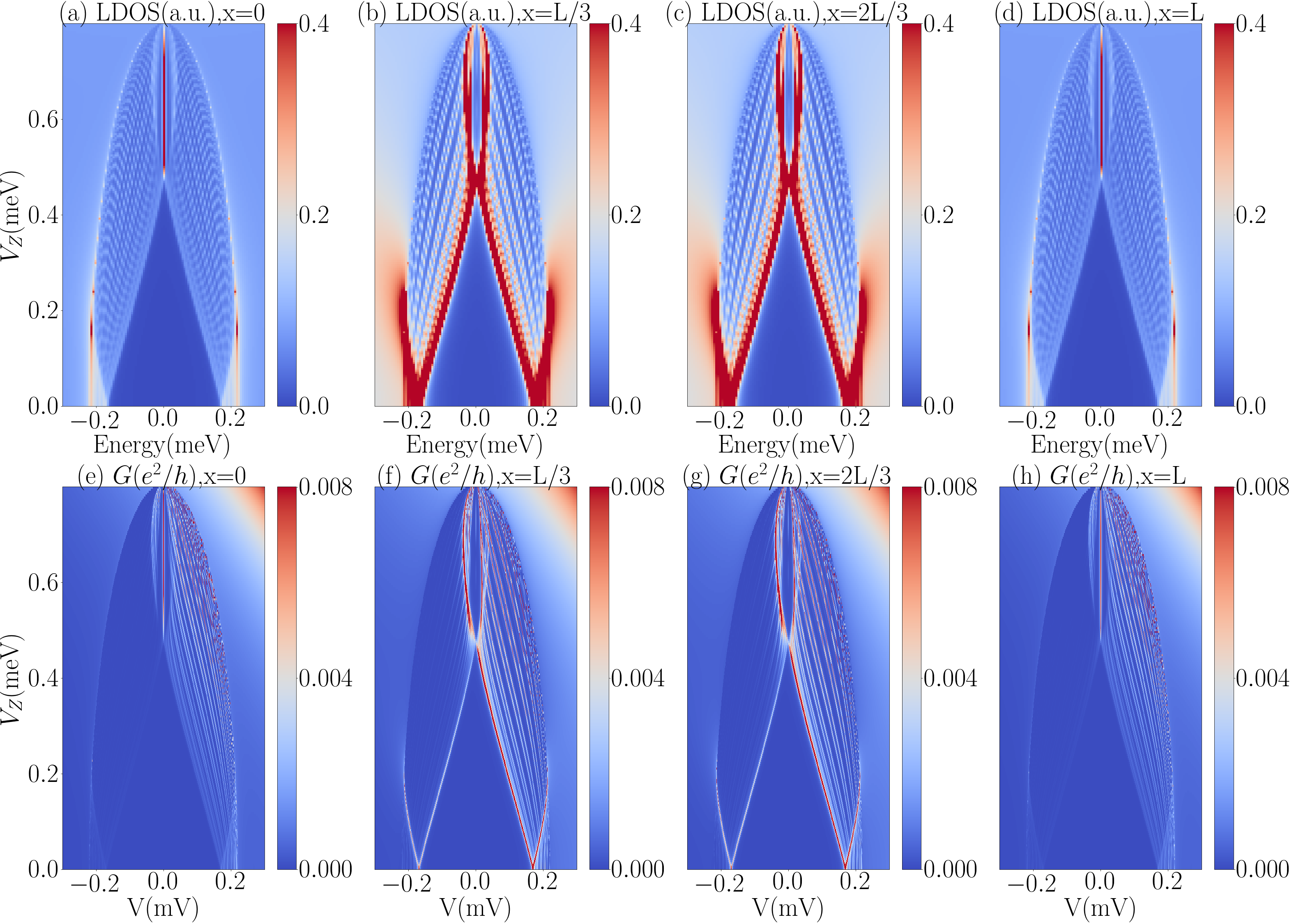}
\caption{(Color online) Local density of states (upper panels) and the conductances (lower panels) as a function of Zeeman field at positions $x=0, L/3, 2L/3, L$ from left to right along the nanowire attached to four barriers (shown in Fig.~\ref{fig:schematic1}).
 The parameters are chosen as the follows: $L=2.5~\mu$m, $L_b=0.1~\mu$m, $\mu=0$~meV, $\Delta_0=0.22$~meV, $\lambda=0.47$~meV, $V_{Z}^*=0.8$~meV, $V_b=1$~meV, $\eta=0.003$~meV, $\Gamma=0.003$~meV, $a=0.01~\mu$m, $t=9.6$~meV, $\alpha=0.25$~meV, $t_b=6$~meV, $t'=3$~meV, and $\alpha'=0.125$~meV.}
\label{fig:LG12}
\end{figure*}

\begin{figure}
\includegraphics[width=0.8\columnwidth]{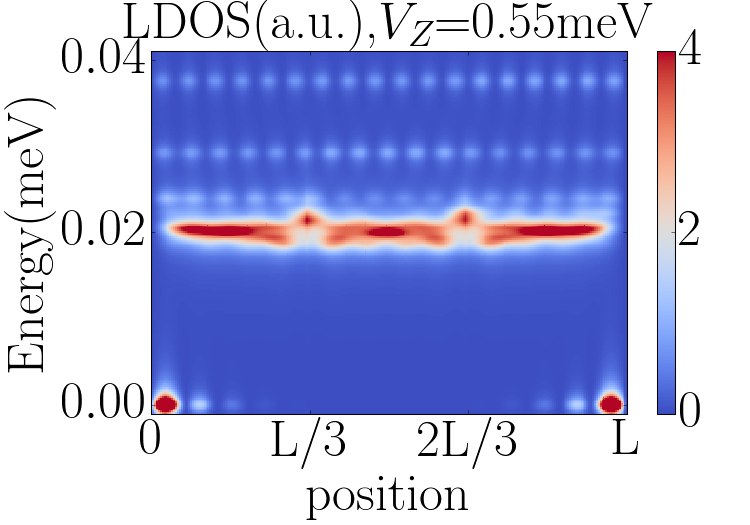}
\caption{(Color online) Spatial profiles of LDOS of the nanowire at $V_Z=0.55$~meV in the topological region($V_Z>V_{Zc}=0.47$~meV). The imaginary term is chosen to be $\eta=0.001$~meV, and the other parameters are the same as those of \ref{fig:LG12}.}
\label{fig:LDOSx12}
\end{figure}

\subsection{Finite chemical potential}\label{sec:finitemu}
In this section, we consider a finite chemical potential case ($\mu=0.15$~meV). For $a=0.01~\mu$m, the values of hopping amplitude $t$ and spin-orbit coupling $\alpha$ in the nanowire are chosen to agree with the parameters claimed in the experimental paper of Grivnin \textit{et al.} ~\cite{Grivnin2018concomitant} ($m^*=0.023m_e$ and $\alpha=0.5$~meV). The parameters of SC are chosen to be $\Delta_0=0.2$~meV and $\lambda=0.3$~meV, which give $\Delta_{ind}=0.12$~meV. The parameters of the barriers and their coupling to the nanowire are chosen to be $V_b=1$~meV, $t_b=10$~meV, $t'=5$~meV, and $\alpha'=0.25$~meV. The barrier parameters are, of course, unknown in the experiment, and we are just making an educated and reasonable guess here.

The calculated LDOS of the nanowire as a function of Zeeman field and bias voltage at $\mu=0.15$~meV are shown in the upper panels of Fig.~\ref{fig:LG11}. The four panels indicate the distinct results at four different positions of the wire where the four barriers are connected to the wire.
 Figure~\ref{fig:LG11}(a) shows the LDOS calculated at one end of the wire ($x=0$). When the Zeeman field increases and goes through the TQPT point at $V_{Zc}=0.34$~meV, the SC gap completely closes and reopens. Above the TQPT point, there is a clearly visible zero-energy peak. This zero-energy peak is the well-known result arising from the Majorana bound state at the end of nanowire~\cite{Sau2010Generic, Lutchyn2010Majorana, Oreg2010Helical}.
  In contrast, as shown in Fig.~\ref{fig:LG11}(b), the LDOS in the bulk of the wire ($x=L/3$) has a zero-energy peak much weaker than that in the end-of-wire LDOS.  We note that this small zero-energy peak is above the TQPT point and thus it could either arise from topological Majorana bound state (MBS) or trivial Andreev bound state. One possibility is the leakage of the MBS wave function localized at the two ends of nanowire with long coherence length, while the other is the leakage of ABS localized in the barriers. This issue will be further discussed at the end of this section. The LDOS on the right side of the wire at $x=2L/3$, which is plotted in Fig.~\ref{fig:LG11}(c), shows similar features as those in Fig.~\ref{fig:LG11}(b), including the small zero-energy peak. By contrast, at $x=L$, Fig.~\ref{fig:LG11}(d) shows the zero-energy peak associated with the Majorana bound state in the opposite end of the wire ($x=L$). This is easy to understand since the system is mirror-symmetrical with respect to the mirror plane at $x=L/2$.

 Next, the corresponding tunneling conductances are presented in the lower panels of Fig.~\ref{fig:LG11}. The four panels show the conductance through one of the four leads coupled to the barrier-nanowire hybrid structure, respectively.
 As a general remark, we note that the conductance through a normal lead in the limit of point contact tunneling qualitatively relates to the local density of states at the same position, in agreement with the general expectations~\cite{PhysRevB.84.155414}. The particle-hole asymmetry in the bulk continuum of conductance results from the nonzero dissipation $\Gamma$. Figure~\ref{fig:LG11}(e) shows a zero-bias conductance peak with a value close to $0.008e^2/h$ at the end of the wire ($x=0$).
 This ZBCP corresponds to the zero-energy peak in Fig.~\ref{fig:LG11}(a) and thus characterizes the Majorana bound state. The fact that the peak value is below the quantized value $2e^2/h$ is a result of the large dissipation relative to the tunneling amplitude in the model. In the presence of the large dissipation, the exact value of the zero-bias conductance peak depends on details of the system such as the height of the barriers and the hopping amplitude of the weak links (and finite temperature, which we neglect in our theory).
 Similar to the corresponding LDOS in Fig.~\ref{fig:LG11}(b), the conductance Fig.~\ref{fig:LG11}(f) has a zero-bias conductance peak with a low value about $0.04e^2/h$, which is lower than the ZBCP in Fig.~\ref{fig:LG11}(e). Again, Figs.~\ref{fig:LG11}(g) and \ref{fig:LG11}(f) [Figs.~\ref{fig:LG11}(h) and \ref{fig:LG11}(e)] look identical.

Finally, to understand the origin of the small zero-energy peak of LDOS above TQPT in the bulk of the nanowire and the corresponding ZBCP in differential conductance, we investigate the spatial distribution of the LDOS at low energies in the whole wire. As shown in Fig.~\ref{fig:LDOSx11}, we present the low-energy LDOS at a fixed Zeeman field above TQPT ($V_Z=0.55$~meV$ > V_{Zc}=0.34$~meV). To improve the visibility of the key features, we use a smaller broadening $\eta=0.001$~meV. At zero energy, we can see that there are two strong peaks near the two ends of the nanowire ($x=0,L$). Interestingly, we note that there are zero-energy peaks at the positions away from the two ends of the nanowire. The height of these zero-energy peaks decay with increasing distance from the wire ends but is still comparable to the peaks at high energies. This zero-energy oscillation pattern can be explained by the presence of Majorana zero modes with long coherence length in a short wire. These Majorana end-mode-induced zero-energy oscillations also explain the appearance of the zero-energy peak of LDOS and the ZBCP of conductance in the bulk of wire in Fig.~\ref{fig:LG11}.

\subsection{Vanishing chemical potential and short coherence length}\label{sec:zeromu}
 In this section, we investigate whether zero-energy peaks in the bulk of the wire exist, in the case that the coherence length is small. To get a system with a short coherence length, the Fermi velocity should be small and the induced SC gap should be large. Here, we choose the hopping and induced SC gap in the nanowire to be $t=9.6$~meV and $\Delta_{ind}=0.15$~meV. In addition, to fix the ratio between the magnetic fields at which the gap closes and superconductivity destroys, we consider a zero-chemical-potential case and choose $V_Z^*=0.8$~meV. The other parameters are adjusted accordingly (see the caption of Fig.~\ref{fig:LG12}) .

The LDOS for the barrier-nanowire structure with the new choice of parameters are shown in the upper panels of Fig.~\ref{fig:LG12}.
If we compare Figs.~\ref{fig:LG12}(a) and \ref{fig:LG12}(d), and Figs.~\ref{fig:LG11}(a) and \ref{fig:LG11}(d), we find that zero-energy peaks in the LDOS at the wire ends ($x=0$ or $L$) look exactly the same.
However, if we look at Figs.~\ref{fig:LG12}(b) and \ref{fig:LG12}(c), there is no evidence for zero-energy peaks in the bulk of the nanowire ($x=L/3$ or $2L/3$) which were found in Figs.~\ref{fig:LG11}(b) and \ref{fig:LG11}(c).
Thus, while the zero-energy peaks at the ends of wire are independent of the choice of parameters, those in the bulk of the wire are not.

We also show the corresponding tunneling conductance in the lower panels of Fig.~\ref{fig:LG12}. By carefully examining the conductance through the bulk of wire in Figs.~\ref{fig:LG12}(f) and \ref{fig:LG12}(g), we can find a small signature of ZBCP whose value is close to zero.  This indicates that the zero-energy peak could exist in the bulk of the wire, although it may not be manifestly observable.

Now we turn to the LDOS profile of whole wire shown in Fig.~\ref{fig:LDOSx12}. Comparing Fig.~\ref{fig:LDOSx11} to Fig.~\ref{fig:LDOSx12}, we note that the zero-energy peaks at the wire ends are stronger in Fig.~\ref{fig:LDOSx12} (compared with Fig.~\ref{fig:LDOSx12}) due to the more localized MZM in Fig.~\ref{fig:LDOSx12}. At the same time, the zero-energy peaks in the bulk of wire decay very fast in Fig.~\ref{fig:LDOSx12}. In particular, the peaks are already invisible at $x=L/3$ or $x=2L/3$ in Fig.~\ref{fig:LDOSx12}. This weak Majorana oscillatory behavior is due to the short coherence length.

\section{Effective potential model}\label{sec:model2}
In the Grivnin \textit{et al}. experiment~\cite{Grivnin2018concomitant}, the width of the tunnel probe is of the same order of magnitude as the wire length. It can be expected that the presence of metallic leads will change the electric charge distribution in the nanowire. Such a change of electrostatic potential can give rise to Andreev bound states~\cite{Huang2018Metamorphosis}. In order to investigate the possible Andreev bound-state contributions to the local properties of nanowire with contacts, we must study contact potential inhomogeneity inside the nanowire introduced by the presence of the leads (which was ignored in the model of Sec.~\ref{sec:model1}) . 
In this section, we therefore consider a Majorana nanowire subject to confinement potential in four contact regions.

\begin{figure}
\raggedleft
\includegraphics[width=1.0\columnwidth]{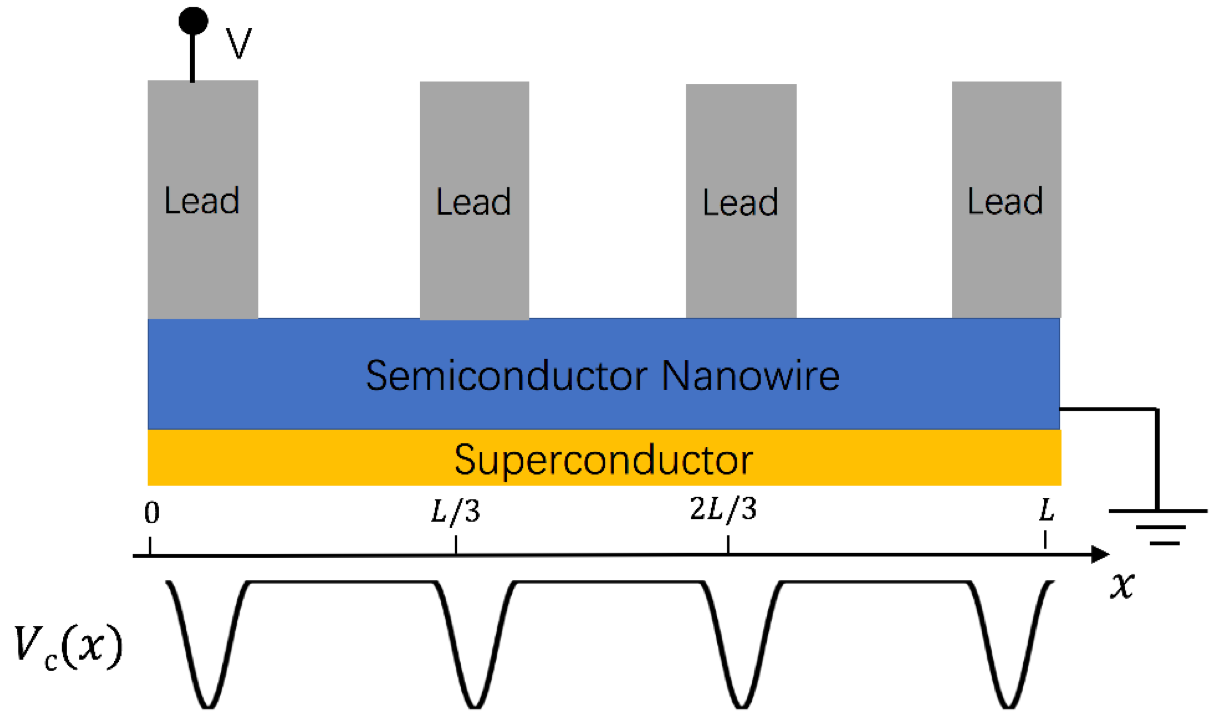}
\caption{(Color online) Schematic for the system composed of a lead and nanowire-superconductor structure, representing the actual setup in Grivnin \textit{et al}. experiment~\cite{Grivnin2018concomitant}.  The nanowire is subject to a confinement potential induced by the leads.}
\label{fig:schematic2}
\end{figure}

A schematic plot of the model in measuring tunneling conductance is shown in Fig.~\ref{fig:schematic2} (to be contrasted with the model shown in Fig.~\ref{fig:schematic1} used in Sec.~\ref{sec:model1}). A semiconductor nanowire with spin-orbit coupling in proximity to an $s$-wave superconductor is attached to four even-spaced leads. The four parts of the nanowire in the contact regions with width $L_c$ are subjected to confinement potential $V_c(x)$.  We assume that the form of potential in one of the contact regions does not depend on whether the lead at this region is active or not.
Note that there are no long potential barriers in every interface between the leads and the nanowire, which is one of the main differences from the contact barrier model discussed in Sec.~\ref{sec:model1}.

With an additional onsite confinement energy $V_c(x)$, the BdG Hamiltonian is $\hat{H}=\frac{1}{2}\int dx\hat{\Psi}^\dag(x)H_{NW}\hat{\Psi}(x)$, where $\hat{\Psi}=(\hat{\psi}_\uparrow,\hat{\psi}_\downarrow,\hat{\psi}^\dag_\downarrow,-\hat{\psi}^\dag_\uparrow)$ are Nambu spinors and
\begin{equation}
\label{eq:H2}
H_{2}=\left(-\frac{\hbar^2}{2m^*}\partial_x^2-i\alpha_R\partial_x\sigma_y-\mu+V_c(x)\right)\tau_z
    +V_Z\sigma_x+\Sigma(\omega,V_Z).
\end{equation}
The smooth contact potential is chosen to be a well-like form in the contact region and zero outside, i.e.,
\begin{equation}
V_c(x)\sim
\begin{cases}
 0& \text{x$\in$ contact-free region},\\
-\frac{V_c}{2}\cos{[2(x-x_c)\pi/L_c]}-\frac{V_c}{2} & \text{x$\in$ contact region},
\end{cases}
\end{equation}
with $V_c$ being the potential well depth, $L_c$ being the width of the contact regions, and $x_c$ being the center of the corresponding contact regions.

We can also write down the Hamiltonian [Eq.~\eqref{eq:H2}] in a tight-binding form as
\begin{align}\label{eq:H2t}
H_2'&=\sum_{j} \Big[-tc^\dag_{j+\delta_x}c_j+(V_c(j)-\mu+2t)c^\dag_{j}c_{j} \\ \nonumber
&+i\alpha(c^\dag_{j+\delta_x}\hat{\sigma}_yc_j-\text{H.c.})+V_Zc^\dag_{j}\hat{\sigma}_xc_j+\Sigma(\omega,V_Z)c^\dag_{j}c_{j}\Big],\nonumber
\end{align}
with the contact potential $V_c(j)$ in a discretized form. In the contact region, $V_c(j)=-\frac{V_c}{2}\cos{[2(j-j_c)\pi/N_c]}-\frac{V_c}{2}$, with $N_c=L_c/a$ and $j_c=x_c/a$; in the contact-free region, $V_c(j)=0$.

\begin{figure*}[t!]
\raggedleft
\includegraphics[width=1.0\textwidth]{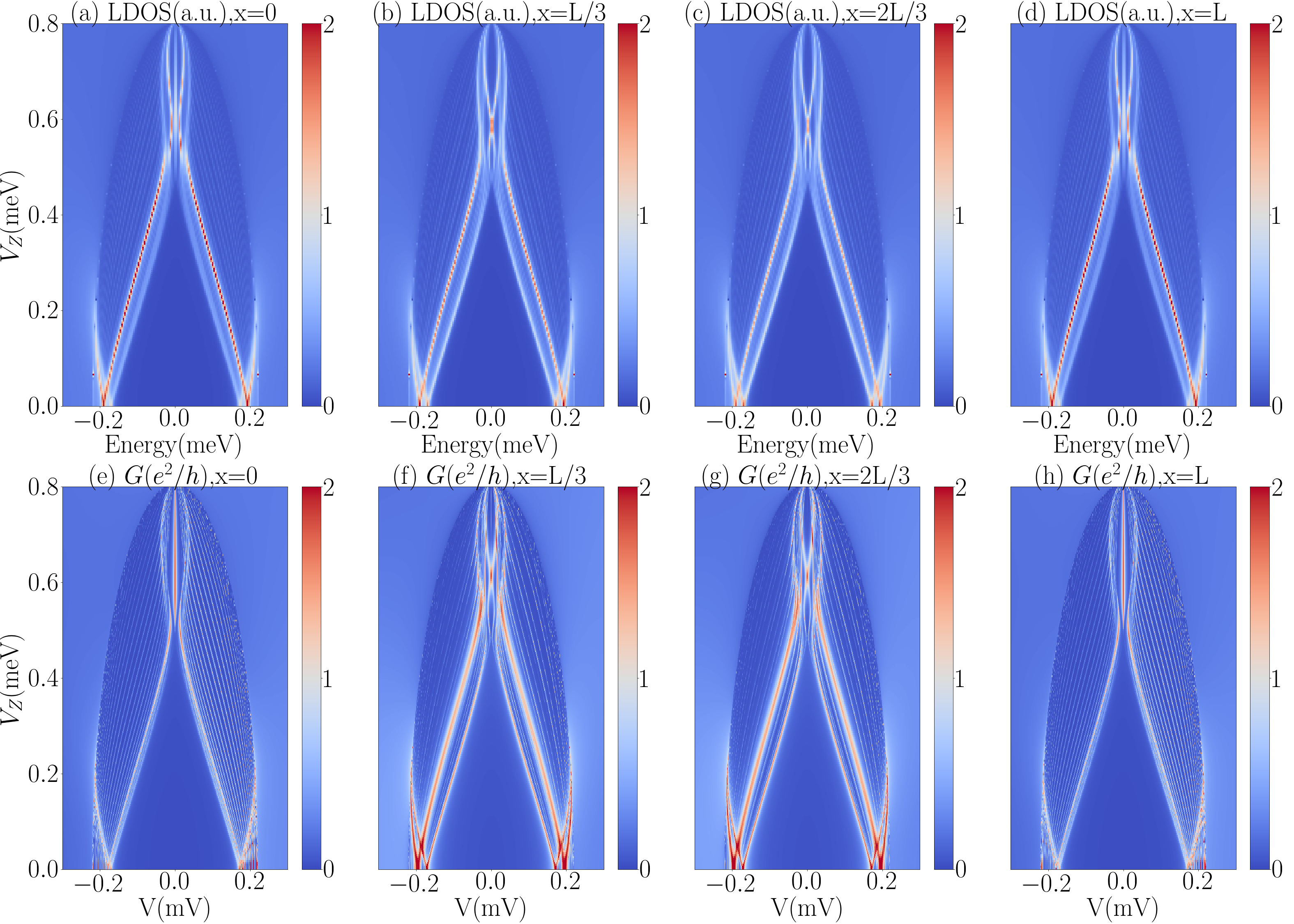}
\caption{(Color online) Averaged local density of states for the four contact regions of the contact potential nanowire as a function of energy and Zeeman field. For convenience, we label the contact region by one of the points inside the region (e.g., $x=0$ is for the region centered at $x_c=L_c/2$; $x=L$ is for the region centered at $x_c=L-L_c/2$). The parameters are chosen as $L=2.5~\mu$m, $L_c=0.25~\mu$m, $V_c=0.65$~meV, $\mu=0$~meV, $\Delta_0=0.22$~meV, $\lambda=0.47$~meV, $V_{Z}^*=0.8$~meV, $V_b=10$~meV, $\eta=0.003$~meV, $\Gamma=0.001$~meV, $a=0.01~\mu$m, $t=9.6$~meV, and $\alpha=0.25$~meV.
}
\label{fig:LG2}
\end{figure*}
\begin{figure}
\includegraphics[width=1.0\columnwidth]{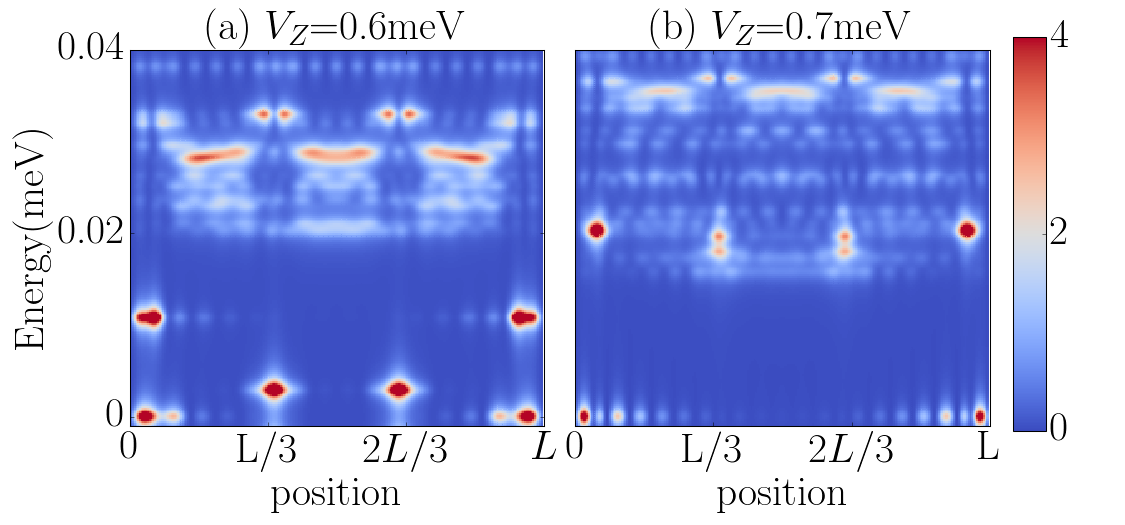}
\caption{(Color online) Spatial profiles of LDOS of the nanowire at (a) $V_Z=0.6$~meV and (b) $V_Z=0.7$~meV in the topological region ($V_Z>V_{Zc}$). The dissipation is chosen to be $\eta=0.001$~meV, and other parameters are the same as those of \ref{fig:LG11}. }
\label{fig:LDOSx2}
\end{figure}

For this nanowire model with position-dependent contact potential, instead of calculating the LDOS at a specific point in the wire defined on the atomic scale, we calculate the average LDOS for every contact region. The average LDOS for a contact region centered at $j_c$ is defined as
\begin{equation}
\overline{\rho}(j_c,\omega)=\frac{1}{N_c}\sum_{|j-j_c|\leq N_c/2}\rho(j,\omega),
\end{equation}
with the sum over all the sites in the contact region.

To calculate the tunneling conductance through one of the four leads to the nanowire, we have to add a square barrier with a height of $V_b=10$~meV in the contact region of the active lead. Different from the four long barriers considered in the multibarrier model of Sec.~\ref{sec:model1}, the barrier we consider here is very short ($L_b=0.02~\mu$m). Note that this barrier at the interface between the active lead and nanowire is not needed in the calculation of LDOS.

For the results presented in this section, we use the same set of parameter values as that in Sec.~\ref{sec:zeromu} and choose $V_c=0.65$~meV and $L_c=0.25~\mu$m for the contact potential. We select this set of parameters because the MBS tail for short coherence length is not our focus in this section.

The averaged LDOS in the four contact regions of the nanowire subjected to the contact potential is shown in the upper panels of Fig.~\ref{fig:LG2}.  Figures~\ref{fig:LG2}(a) and \ref{fig:LG2}(d) show the averaged LDOS in the two contact regions near the two ends of the nanowire, respectively. Above the critical Zeeman field $V_{Zc} =0.47$~meV, besides a zero-energy peak, there is a pair of peaks at finite energy below the gap. The two peaks come together at around $V_Z=0.6$~meV and then split at a somewhat higher Zeeman field, but they come together again at around $V_Z^*=0.8$~meV. Note that this pair of near-zero-energy peaks is absent in Figs.~\ref{fig:LG12}(a) and ~\ref{fig:LG12}(d). Thus, we expect that this pair of near-zero-energy peaks is associated with ABS induced by confinement potential. This expectation is supported by the fact that the ABS peak behavior changes with varying confinement potential. In Figs.~\ref{fig:LG2}(b) and \ref{fig:LG2}(c), we find a pair of near-zero-energy peaks with the same behavior in the averaged LDOS in the two contact regions near the middle of the nanowire ($x=L/3,2L/3$).

The calculated tunneling conductance through the nanowire from probing leads at four different positions in the four contact regions is shown in the lower panels of Fig.~\ref{fig:LG2}. The tunneling conductance calculated with the probing lead near
two ends of nanowire [in Figs.~\ref{fig:LG2}(e) and \ref{fig:LG2}(h)] manifests zero-bias coherence peaks within the topological regime with a height of $2e^2/h$. They are robust to contact potential and have higher values than those of the ZBCP in Figs.~\ref{fig:LG12}(e) and \ref{fig:LG12}(h). Meanwhile, a pair of near-zero-bias conductance peaks appears in the conductance when the probing lead is in the bulk of nanowire ($x=L/3, 2L/3$).  This pair of conductance peaks arises from Andreev bound states at the energy of the peaks shown in Figs.~\ref{fig:LG2}(b) and \ref{fig:LG2}(c). These are nontopological quasi-MZM peaks induced by the tunnel probes themselves.

To further investigate the contact-potential-induced Andreev bound state in the system, we turn to the spatial profile of LDOS. The LDOS for the contact-potential nanowire at fixed $V_Z=0.6$~meV and $V_Z=0.7$~meV is shown in Fig.~\ref{fig:LDOSx2}.
 At $V_Z=0.6$~meV [Fig.~\ref{fig:LDOSx2}(a)], there are six peaks below the energy at $0.02$~meV. While four of the peaks are at the wire ends, the other two near-zero-energy peaks are in the bulk of the wire. Strictly speaking, these two peaks are at $x=L/3$ and $x=2L/3$, the centers of the two contact regions. Thus, the position of these two strong peaks in the LDOS confirm their origin from the contact potential. Similarly, the origin of the four peaks near the two wire ends can also be determined. We believe that two of them arise from ABS in the contact regions, while the other two are from MBS. To check this, in Fig.~\ref{fig:LDOSx2}(b) we increase the Zeeman field to $V_Z=0.7$~meV, resulting in four peaks in the four contact regions going to higher energies and two peaks staying at zero energy. This is consistent with the prediction that the four contact regions along the wire host four ABS in this contact potential model, showing the extent to which the details of the probe matter.

\section{Conclusions}\label{sec:conclusion}
We have investigated the low-energy physics of a semiconductor-superconductor hybrid structure coupled to multiple tunnel probes along the length of the wire using the numerical solution of a tight-binding model. The proximity effects induced by the superconductor are incorporated through a self-energy interface contribution to the  nanowire Green's function, which allows us to account for both the induced superconducting pair potential and the proximity-induced low-energy renormalization. The effect of the tunnel probes is considered at two levels using two complementary models. First, we simply couple the probes to the wire through a barrier region modeled as a finite width potential plus a narrow weak link. The second model takes into account the fact that the very presence of the (metallic) contacts alters the local electrostatic potential inside the wire. This effect is incorporated as a position-dependent electrostatic potential. Which model is more appropriate for a specific experimental system depends on a lot of unknown details, but it is likely that the experimental systems are a combination of both.

We find that, generically, the presence of the tunnel contacts perturbs the system locally and induces low-energy Andreev bound states localized near the bulk contacts. By contrast, the perturbation associated with the end-of-wire contacts does not affect the emergence of Majorana zero modes in the topological SC regime. Furthermore, as expected, the emergence of the MZM-induced ZBCPs at the opposite ends of the wire is correlated, even when the local conditions are slightly different. We find that the dominant features measured by the bulk probes are generally associated with contact-induced ABSs, rather than genuine bulk states. However, if the contact-induced perturbation is weak enough, the induced ABS crosses zero energy slightly above the TQPT, producing a gap closing-and-reopening feature that mimics the expected behavior of the bulk gap. Nonetheless, the exact value of the ABS gap depends on the local conditions and, in general, is expected to be different for different bulk probes, in contrast with the expected position-independent value of a genuine bulk topological gap. In addition,  for stronger contact-induced perturbations the contact-induced  ABS may cross zero energy multiple times and produce ZBCP-like features in the bulk. In this case, the correlation between the emergence of the MZM-induced ZBCP at the ends and the reopening of the quasiparticle gap in the bulk is lost.

Based on our analysis, we conclude that multiprobe measurements represent a significant step forward as compared to the ``standard'' end-of-wire tunneling measurement.  In fact, we strongly urge all nanowire experimental groups to follow the lead of Ref.~[\onlinecite{Grivnin2018concomitant}] and carry out detailed multiprobe measurements in as many devices as possible. However, the experimental results have to be interpreted cautiously, while making efforts to minimize the contact-induced perturbations. We emphasize that creating a large barrier between the wire and the leads (i.e., working in the extreme weak-coupling limit) is not enough to ensure that contact-induced perturbation is negligible, as it contains an electrostatic component that is independent of the barrier strength. More specifically, the very presence of a metallic lead can locally perturb the electrostatic environment. Note that the typical scale associated with variations of the electrostatic potential across the wire is $\sim10^2~$mV, while a strong perturbation may involve local variations of the order of $1~$mV. To minimize these electrostatic effects we propose two possible setups. The first suggestion is to engineer extremely narrow bulk contacts. The rationale behind this geometry is that, on the one hand, the local variation of the electrostatic  potential produced by a narrow contact due to additional screening is weaker and, on the other hand, short-range potential perturbations are less likely to generate low-energy ABSs. The second proposal is to cover the entire bulk of the proximitized nanowire with 3--5 wide contacts separated by very narrow gaps. In addition to ensuring a uniform electrostatic environment, this geometry addresses the issue of tunneling into bulk states, which do not couple well to local probes and, consequently, produce very weak signatures (as compared with the bound states localized near the contacts). In addition to addressing the engineering problems, self-consistent 3D Schrodinger-Poisson calculations should be used as a theoretical tool for assessing the feasibility of these proposals. We are optimistic that future experiments along the line of Ref.~[\onlinecite{Grivnin2018concomitant}] will lead to definitive evidence supporting the existence of topological MZMs in Majorana nanowires, even if the first measurements are likely to be fraught with the invasive nature of the tunnel probes producing bulk Andreev bound states.

\begin{acknowledgements}
The authors are indebted to C.-X. Liu, H. Pan, and W. S. Cole for discussions. The authors are particularly grateful to M. Heiblum for useful correspondence on the experimental details of Ref.~[\onlinecite{Grivnin2018concomitant}]. This work is supported by the Laboratory for Physical Sciences and Microsoft. Y.H. acknowledges funding from the China Scholarship Council. We also acknowledge the University of Maryland supercomputing resources made available in conducting the research reported in this paper.
\end{acknowledgements}

\bibliography{BibMajorana}
\end{document}